\documentclass[aps,prd,twocolumn,showpacs,groupedaddress,floatfix]{revtex4}

\usepackage{graphicx}
\usepackage{longtable}
\DeclareGraphicsExtensions{.eps, .eps, .jpg, .png}

\begin{document}

\preprint{hep-th/0501028\\ January 2005}
\date{\today}

\title{Brane-Induced Gravity's Shocks}

\author{Nemanja Kaloper\footnote{\tt kaloper@physics.ucdavis.edu} }
\affiliation{Department of Physics, University of California,
Davis, CA 95616 }

\begin{abstract}
We construct exact gravitational field solutions for a
relativistic particle localized on a tensional brane in
brane-induced gravity. They are a generalization of gravitational
shock waves in 4D de Sitter space. We provide the metrics for both
the normal branch and the self-inflating branch DGP braneworlds,
and compare them to the 4D Einstein gravity solution and to the
case when gravity resides only in the 5D bulk, without any
brane-localized curvature terms. At short distances the wave
profile looks the same as in four dimensions. The corrections
appear only far from the source, where they differ from the long
distance corrections in 4D de Sitter space. We also discover a new
non-perturbative channel for energy emission into the bulk from
the self-inflating branch, when gravity is modified at the de
Sitter radius.
\end{abstract}
\pacs{11.25.-w, 11.25.Mj, 98.80.Cq, 98.80.Qc \hfill
hep-th/0501028}

\maketitle


In brane-induced gravity \cite{dgp} gravitational force between
masses on the brane should arise predominantly from the exchange
of an unstable bulk graviton resonance, which falls apart at very
large scales, instead of a stable zero-mode graviton governed by
the usual four-dimensional (4D) General Relativity (GR). Far from
sources the law of gravity is modified, and changes from 4D to a
fully higher-dimensional one. Subtleties of gravitational
perturbation theory in DGP have been explored in
\cite{strongcouplings,lpr,giga,nira} with the particular attention
on the couplings of the longitudinal gravitons. It has been
suggested that the brane extrinsic curvature may play the role of
a coupling controller: the extrinsic curvature, responding to a
source mass on the brane, may help to tame perturbation theory
\cite{strongcouplings,nira}. Exploring this interplay between the
curvature and the graviton couplings on exact solutions of DGP
field equations could clarify the status of effective 4D theory.
The DGP equations are, however, notoriously difficult to solve for
localized sources \cite{schwarz}.

In this Letter we present the first example of an exact solution
for a localized particle in DGP models. We derive the
gravitational field for a relativistic particle on a tensional
brane in 5D. It generalizes 4D gravitational shock waves
\cite{aichsexl,thooft,gabriele} in de Sitter space
\cite{hota,pogri,kostas}. We present the closed form metrics for
both the normal branch and the self-inflating branch DGP
braneworlds \cite{dsbranes}. The scalar graviton is absent in
these metrics. One expects this in weakly-coupled perturbation
theory, where the scalar graviton decouples from
ultra-relativistic particles since to leading order its source
vanishes, $T^{\mu}{}_\mu = 0$, and the corrections are suppressed
by the source mass to momentum ratio $m/p \ll 1$ (e.g., hence the
gravitational shock wave of 4D GR is also a solution in
Brans-Dicke theory). Our results show that this persists in DGP,
and that the dangerous strongly coupled mode found in perturbation
theory does not destroy the solutions at large momenta after all.
Further, we find that as long as the brane-localized terms are
present, at short distances the wave profile behaves exactly like
in 4D GR \cite{aichsexl,thooft}. The deviations appear only far
from the source, where they differ from the 4D GR corrections in
de Sitter space. We also suggest a new way for checking graviton
effective field theory with our shock waves.

DGP braneworlds \cite{dgp} are given by a bulk action with metric
kinetic terms in both the bulk and on the brane, with couplings
set by bulk and brane Planck scales, $\kappa^2_4 = 1/M^2_4 $ and $
\kappa^2_5 = 1/M^3_5$. In 5D, the field equations are \cite{dgp}
\begin{equation}
M^3_5 G_5^A{}_B  + M^2_4 G_4^\mu{}_\nu \delta^A_\mu \delta^\nu_B
\delta(w) = - T^\mu{}_\nu  \delta^A_\mu \delta^\nu_B \delta(w)\, ,
\label{fieldeqs}
\end{equation}
where $G_5^A{}_B$, $G_4^\mu{}_\nu$ are bulk and 3-brane-localized
Einstein tensors; $T^{\mu}{}_\nu$ is the brane stress-energy;
$\{A,B\}$ and $\{\mu,\nu\}$ are bulk and 3-brane world-volume
indices, respectively. We use Gaussian normal coordinates for the
5D metric, $ds_5^2 = dw^2 + g_{\mu\nu}(w,x) dx^\mu dx^\nu$,
orbifolding by a $Z_2$ symmetry $ w \rightarrow - w$ around the
brane at $w=0$. For a brane with tension $\lambda \ne 0$ and no
matter, $T^{\mu}{}_{\nu} = - \lambda \delta^\mu{}_\nu$. The
solution is a de Sitter 3-brane in a flat bulk \cite{dsbranes},
\begin{equation}
ds^2_5 = (1- \epsilon H|w|)^2 ds^2_{4dS} + dw^2 \, .
\label{background}
\end{equation}
Apart from $\epsilon = \pm 1$ in (\ref{background}), this is
identical to the 5D version \cite{kalinde} of the inflating VIS
domain wall \cite{vis}. We shall use $ds^2_{4dS} = -(1-H^2r^2)
dt^2 + \frac{dr^2}{(1 - H^2 r^2)} + r^2 d\Omega_2$, the static de
Sitter metric instead of the spatially flat one employed in
\cite{kalinde}. The space-time of (\ref{background}) is a 4D de
Sitter hyperboloid in a 5D Minkowski bulk. Here $\epsilon$ arises
because we can retain either the interior $\epsilon = + 1$ or the
exterior $\epsilon = -1$ of the hyperboloid after orbifolding,
thanks to the brane curvature. The junction conditions relate the
4D curvature and the brane tension as \cite{dsbranes}
\begin{equation}
H^2 + \epsilon \frac{2 M^3_5}{M_4^2} H = \frac{\lambda}{3 M^2_4}
\, . \label{hubble}
\end{equation}
The solutions with $H >0$ differ for each of $\epsilon = \pm 1$
($H<0$ cases are their $PT$ transforms). On the normal branch,
$\epsilon = +1$, the solution reduces to the 4D Friedman equation
in the limit $M_5 \ll M_4$, $3H^2 \simeq \lambda/M^2_{4}$. In the
bulk, we keep the interior of the hyperboloid. It has finite
volume and so the perturbative 4D graviton exists. The brane
curvature terms  $\sim M^2_{4} \int d^4x \sqrt{g_4} R_4/2$
suppress the couplings of the $m_g > 0$ graviton KK modes,
producing the 4D effective theory when $M_5 \ll M_4$. On the
self-inflating branch with $\epsilon = -1$, at low tensions
$\lambda \ll  12 M^6_5/M^2_4$, the eq. (\ref{hubble}) yields $H
\sim 2 M^3_5/M_4^2$. In the bulk we keep the exterior of the
hyperboloid which has infinite bulk volume and so there is no
perturbative 4D graviton. The effective theory arises only from
the exchange of the bulk resonance.

Suppose now there is also a single relativistic particle, say a
photon, with a momentum $p$ on the brane, moving along some null
geodesic of the brane-induced metric. Its stress-energy tensor
sources an additional gravitational field. To find it, one could
boost the linearized gravitational field for a massive particle by
an infinite amount, simultaneously taking the limit $m \rightarrow
0$ such that $m \cosh \gamma = p$ stays finite
\cite{aichsexl,hota,pogri,roberto}. Viewing the metric as an
expansion in the powers $m$, in the relativistic limit the terms
of order higher than $m$ vanish because there is only one factor
of the boost parameter $\cosh \gamma$, and the linearized solution
becomes exact! Alternatively, one can use a variant of a very
elegant cut-and-paste technique developed by Dray and 't Hooft
\cite{thooft} for flat 4D backgrounds, and later applied to
general 4D GR backgrounds by Sfetsos \cite{kostas}. We will follow
this route here, by extending the technique to DGP. First, we cast
(\ref{background}) in suitable null coordinates, defining $u =
\frac{1}{H} \sqrt{\frac{1-Hr}{1+Hr}} \exp(Ht)$ and $v =
\frac{1}{H} \sqrt{\frac{1-Hr}{1+Hr}} \exp(-Ht)$. We also change
$w$ to $|z| = -\frac{1}{\epsilon H} \ln(1 - \epsilon H |w|)$ (note
that $\delta(w) = \delta(z)$), to simplify evaluating the
curvature tensors with a conformal map as in \cite{conformal}. The
metric (\ref{background}) becomes
\begin{equation}
ds_5^2 = e^{-2 \epsilon H |z|} \Bigl\{ \frac{4 dudv}{(1+H^2 uv)^2}
+ (\frac{1-H^2 uv}{1+H^2 uv})^2 \frac{d\Omega_2}{H^2} + dz^2
\Bigr\} \, . \label{nullback}
\end{equation}
Let the photon move along the $v$-axis on the brane, i.e. along
the null geodesic $u=0$. The observer on the North Pole of the
brane de Sitter space sees a photon streaming away along the past
horizon, starting from $r=0$ at infinite past. Using the trick of
Dray and 't Hooft to construct the photon's gravitational field,
we introduce a jump in the $v$ coordinate at $u=0$ \cite{thooft}:
let $v \rightarrow v + \Theta(u) f$ and $dv \rightarrow dv +
\Theta(u) df$ where $f$ is the wave profile which depends only on
the spatial transverse coordinates (in our case, the angles on the
2-sphere and $z$), and $\Theta(u)$ is just the usual step
function. Changing the coordinates to $\hat v = v + \Theta(u) f$
yields $dv \rightarrow d \hat v - \delta(u) f du$. Substituting
$v,dv \rightarrow \hat v, d\hat v - \delta(u) fdu$ in (4) and
dropping the carets gives
\begin{eqnarray}
ds_5^2 &=& e^{-2 \epsilon H |z|} \Bigl\{ \frac{4 dudv}{(1+H^2
uv)^2} - \frac{4 \delta(u) f du^2}{(1+H^2 uv)^2} + \nonumber \\
&& ~~~~~~~~~~~~~~ + (\frac{1-H^2 uv}{1+H^2 uv})^2
\frac{d\Omega_2}{H^2} + dz^2 \Bigr\} \, . \label{wave}
\end{eqnarray}
Substituting (\ref{wave}) into (\ref{fieldeqs}), including the
photon source in $T^{\mu}{}_{\nu} = - \lambda \delta^\mu{}_\nu + 2
(p/\sqrt{g}) g_{uv} \delta(\theta) \delta(\phi) \delta(u)
\delta^\mu_v \delta^u_\nu $, and evaluating the curvature, bearing
in mind that $\delta(u)$ and its derivatives are distributions (so
$u \delta(u) = 0$, $u^2 \delta^2(u) = 0$ and $f(u) \delta'(u) = -
f'(u) \delta(u)$ \cite{thooft,kostas}), we find the independent
field equations \cite{kalshocks}. One is just eq. (\ref{hubble}).
The other comes from the $G_{5~uu}$ component of (\ref{fieldeqs})
and yields a {\it linear} field equation for the wave profile:
\begin{eqnarray}
&& \frac{M^3_5}{M_4^2 H^2} \Bigl(\partial_z^2 f - 3 \epsilon H
\partial_{|z|} f + H^2( \Delta_2 f + 2f) \Bigr) + \nonumber \\
&& ~~~~~~~~~~~~~~~ + (\Delta_2 f + 2f) \delta(z) =
\frac{2p}{M^2_4} \delta(\Omega) \delta(z) \, . \label{feqn}
\end{eqnarray}
Here $\Delta_2$ and $\delta(\Omega)$ are the Laplacian and the
$\delta$-function on a 2-sphere, peaked at $\theta=0$. In the
limit $M_5 \rightarrow 0$ the bulk derivatives disappear, one can
factor out $\delta(z)$ and recover the 4D de Sitter equation of
\cite{hota,kostas}.

It is simpler to solve eq. (\ref{feqn}) for two sources with the
same momentum $p$, running in the opposite directions in the
static patch (\ref{nullback}) \cite{hota,pogri}. In this way we
avoid some unphysical divergences when the last term in
(\ref{modesn}) vanishes for $l=1$, which cancel out anyway by
symmetry. The two-source solution correctly represents the limit
of infinite boost of the Schwarzschild-de Sitter geometry
\cite{hota,pogri}. We add an extra term, $\frac{2p}{M^2_4}
\delta(\Omega') \delta(z)$, on the RHS of (\ref{feqn}) where
$\delta(\Omega')$ is peaked at $\theta = \pi$. To return to a
single source, we can take the solution and multiply it by
$\Theta(\pi/2 - \theta)$ as in \cite{kostas}. So one particle
moves along $\theta = 0$ and the other along $\theta = \pi$. With
this choice of trajectories we use the addition theorem for
spherical harmonics to replace them with Legendre polynomials
$P_l(\cos \theta)$ in the expansion for the wave profile: $f =
\sum_{l=0}^\infty ( f_l^{(+)}(z) P_l(\cos \theta) + f_l^{(-)}(z)
P_l(-\cos \theta) )$. Here $f_l^{(\pm)}(z)$ are the bulk wave
functions; $f_l^{(+)}$ is sourced by the photon at $\theta=0$ and
$f_l^{(-)}$ by the photon at $\theta = \pi$. By orthogonality and
completeness of Legendre polynomials, the modes $f_l^{(\pm)}(z)$
obey the same differential equation,
\begin{eqnarray}
&& \partial_z^2 f_l - 3 \epsilon H
\partial_{|z|} f_l + H^2(2 - l(l+1) ) f_l = \nonumber \\
&& ~~~ = \frac{M_4^2 H^2}{M^3_5} \Bigl( \frac{(2l+1)p}{2 \pi
M^2_4} - (2 - l(l+1))f_l \Bigr) \delta(z) \, . \label{modesn}
\end{eqnarray}
We interpret $\delta$-function on the RHS by pillbox integration
as a matching condition for the first derivatives of $f_l^{(\pm)}$
on the brane.
The remaining boundary conditions come from requiring
orbifold symmetry $f_l(-z) = f_l(z)$ and square integrability in
the bulk, so that the solutions are localized on the brane. Since
both $f_l^{(\pm)}$ solve the same boundary value problem,
$f_l^{(+)} = f_l^{(-)} = f_l$. By (\ref{modesn}), $f_l$'s are
simple exponentials, and to be localized on the brane (i.e.
square-integrable) they must be $\sim e^{- [2 l +(1-3\epsilon)/2]
\, H |z|}$. Because $P_l(-x) = (-1)^l P_l(x)$, the solution will
be an expansion only in even-indexed polynomials $P_{2l}(\cos
\theta)$, like in 4D \cite{hota,kostas}. Solving (\ref{modesn})
for the bulk wave functions $f_l$'s \cite{kalshocks} yields the
solution for two relativistic photons on the brane moving in
opposite directions:
\begin{eqnarray}
f(\Omega,z) &=& - \frac{p}{2\pi M^2_4} \sum_{l=0}^\infty
\frac{4l+1}{(2l - 1 + \frac{(1-\epsilon){\tt g}}{2})
(l+1 + \frac{(1+\epsilon){\tt g}}{4})} \nonumber \\
&& ~~~~~~~~~~~~~ \times e^{- [2 l +(1-3\epsilon)/2] \, H |z|}
P_{2l}(\cos \theta) \, , \label{sols}
\end{eqnarray}
where ${\tt g} = 2 M^3_5/(M^2_4 H) = 1/(H r_c)$ (see \cite{dgp}).
The solution can be checked by direct substitution into
(\ref{feqn}), (\ref{modesn}). When ${\tt g}=0$, at $z=0$ this
reproduces the 4D series solution of \cite{hota} for both
$\epsilon = \pm 1$. We can bring the series (\ref{sols}) to a more
compact form. Factorizing the coefficients of the expansion, and
defining $\tau = e^{- H |z|}$ and $x = \cos \theta$,
\begin{eqnarray}
f(\Omega,z) = - \frac{[3 - (1-\epsilon) {\tt g}] \, p }{(3 +
\epsilon {\tt g})\, \pi M^2_4} \sum_{l=0}^\infty \frac{\tau^{2
l+(1-3\epsilon)/2}}{2l - 1 +
\frac{(1-\epsilon){\tt g}}{2}} P_{2l}(x) && \nonumber \\
- \frac{[3 + (1+\epsilon){\tt g}] \, p}{2 (3 + \epsilon {\tt g})
\pi M^2_4}  \sum_{l=0}^\infty \frac{\tau^{2 l
+(1-3\epsilon)/2}}{l+1 + \frac{(1+\epsilon){\tt g}}{4}} P_{2l}(x)
\,\, . ~~~~ &&  \label{facsols}
\end{eqnarray}

Note that while for $\epsilon = 1$ the series (\ref{facsols}) is
finite, for $\epsilon = -1$, the self-inflating branch solutions,
it displays a spectacular behavior as ${\tt g}\rightarrow 1$, when
the brane tension vanishes, where the $l=0$ term of the first sum
has a pole (${\tt g}=3$ is regular, as seen from (\ref{sols})). A
closer look \cite{kalshocks} shows that in this limit the only
finite $l=0$ solution of (\ref{modesn}) is the delocalized bulk
mode, not present in (\ref{sols}). For ${\tt g}=1$ gravity is
modified at the scale equal to the cosmological horizon on the
brane. Brane and bulk begin to ``resonate": in a time-dependent
problem, as ${\tt g}\rightarrow 1$ slowly, the $l=0$ mode would
grow larger, and at some point it may begin to produce delocalized
gravitons, taking energy from the self-inflating brane into the
bulk. This indicates an onset of a dramatic new instability, which
warrants closer investigation.

Now, recall the generating function for Legendre polynomials, $(1
- 2x \tau + \tau^2)^{-1/2} = \sum_{l=0}^\infty P_l(x) \tau^l$.
Using $\lim_{\varepsilon\rightarrow 0} \int^\tau_\varepsilon d
\vartheta \vartheta^{l+\beta}= \frac{\tau^{l+\beta+1}}{l+\beta+1}
- \lim_{\varepsilon\rightarrow 0} \int^\varepsilon d \vartheta
\vartheta^{l+\beta}$ to regulate divergences when $l + \beta \le
-1$, integrate over $\tau$; after straightforward manipulations
eq. (\ref{facsols}) becomes
\begin{eqnarray}
f(\Omega,z) = \frac{p}{2\pi M^2_4}
\frac{\tau^{\frac{1-3\epsilon}{2}}}{1 - \frac{1-3\epsilon}{4}{\tt
g}} -  \frac{p}{2(3 + \epsilon {\tt g}) \pi M^2_4} \times
~~~~~~~~ && \nonumber \\
\times \int^\tau_0 d\vartheta \Bigl(
\frac{1}{\sqrt{1-2x\vartheta+\vartheta^2}}
+ \frac{1}{\sqrt{1+2x\vartheta+\vartheta^2}} -2 \Bigr)
~~~~~ && \label{intsols}\\
\times \Bigl( \frac{[3 - (1-\epsilon) {\tt g}]
\vartheta^{\frac{{\tt g}(1-\epsilon)-4}{2}}}{\tau^{\frac{({\tt
g}-3)(1-\epsilon)}{2}}} + \frac{[3 + (1+\epsilon){\tt
g}]\vartheta^{\frac{{\tt g}(1+\epsilon)+2}{2}}}{\tau^{\frac{({\tt
g}+3)(1+\epsilon)}{2}}} \Bigr) \, . \nonumber &&
\end{eqnarray}
Since (\ref{intsols}) is a Green's function in the transverse
directions, it has short-distance singularities at $x = \pm 1$,
but is well-behaved elsewhere. We will not attempt to evaluate it
for the general case. Here we only look at the formula for $f$ on
the brane $z=0$ (i.e. $\tau = 1$) at transverse distances ${\cal
R}$ well inside the cosmological horizon, ${\cal R } \ll H^{-1}$.
Note that if we rewrite eq. (\ref{hubble}) as $(1 + \epsilon {\tt
g}) H^2 = \lambda/(3M^2_4)$ we find that on the normal branch
${\tt g}$ can vary between zero (4D) and infinity (5D), with the
``resonance" mentioned above at ${\tt g}=1$. On the self-inflating
branch, ${\tt g} \le 1$ or $r_c \ge 1/H$, as long as $\lambda \ge
0$. In the limit ${\tt g}\rightarrow 1$ on the self-inflating
brane the tension vanishes.

When ${\tt g} = 0$, (\ref{intsols}) is identical to the 4D case on
both branches, as remarked above. Indeed, the integrals give
\begin{equation}
f_{4D}(\Omega) = \frac{p}{2 \pi M^2_4} \Bigl( 2 - x \,
\ln[\frac{1+x}{1-x}] \Bigr) \, . \label{4dsols}
\end{equation}
The metric transverse to the null particle on the brane is
$ds_2^2|_{z=u=0} = d\Omega_2/H^2$ and so the proper ``radial"
distance is measured by the polar angle $\theta$. For small
angles, ${\cal R} \simeq \theta/H$, $x = 1- H^2 {\cal R}^2/2$, and
(\ref{4dsols}) reduces precisely to the flat 4D solution
\cite{aichsexl,thooft}: up to ${\cal O}({\cal R}^2/H^{-2})$
corrections, and with the sign conventions of \cite{hota,kostas},
\begin{equation}
f_{4D}(\Omega) = \frac{p}{\pi M^2_4} + \frac{p}{\pi M^2_4}
\ln(\frac{\cal R}{2H^{-1}}) \, . \label{4dsolflat}
\end{equation}
The constant term in this equation can always be recovered in the
4D flat case by a diffeomorphism. The integrals for ${\cal O}({\tt
g})$ corrections contain terms like $\int d\zeta
\ln(1+b\zeta)/\zeta$ and cannot be written in closed form
\cite{gradrhyz}. However, at short distances the singular ${\cal
O}({\tt g})$ powers in the integrand precisely cancel out. In the
remainder, since the leading order singularity in (\ref{intsols})
is logarithmic around ${\tt g}=0$, the additional logarithms
soften the corrections further, and render them finite as ${\cal
R} \rightarrow 0$. Hence at short distances the solution looks
exactly the same as in 4D.

In fact this persists for any finite value of ${\tt g}$. If we
rewrite the integrals (\ref{intsols}) using the Euler
substitutions $\zeta = \sqrt{1\mp 2x\vartheta+\vartheta^2}+
\vartheta$ and consider the limit $1-x \ll 1$, we extract the
leading singularity. It comes from the term $\sim - \frac{p}{\pi
M^2_4} \int_1^{1+\sqrt{2(1-x)}} \frac{d\zeta}{\zeta-x} =
\frac{p}{\pi M^2_4} \ln\sqrt{\frac{1-x}{2}} + {\rm
finite~terms}\,$ for all values of ${\tt g}$ and $\epsilon$.
Substituting $x = 1- H^2 {\cal R}^2/2$, we recover the leading
logarithm in (\ref{4dsolflat}). The subleading corrections in
general differ from the terms in the expansion of the 4D de Sitter
solution (\ref{4dsols}) around the short distance limit in flat
space (\ref{4dsolflat}). In that case one finds corrections to
come as even powers of $H{\cal R}$. In DGP, however, the
corrections in (\ref{logs}) will come as odd powers of $H{\cal R}$
as well, starting with the linear term, signalling the hidden
fifth dimension. Thus in general we will have
\begin{eqnarray}
f(\Omega) &=& \frac{p}{\pi M^2_4} \Bigl((1 + a_1 H^2 {\cal R}^2 +
...)
\ln(\frac{\cal R}{2H^{-1}}) \nonumber \\
&&~~~~~ + {\rm const} + b_1 H {\cal R} + b_2 H^2 {\cal R}^2 + ...
\Bigr) \, ,
 \label{logs}
\end{eqnarray}
where the coefficients $a_k, b_k$ are numbers that can be computed
explicitly for given values of ${\tt g}$ and $\epsilon$.

To see how 5D gravity reemerges, on the normal branch we can take
the limit $g \rightarrow \infty$. Then the background reduces to
the inflating brane in 5D Minkowski bulk \cite{kalinde}. Using
(\ref{facsols}) to find the contribution of the second sum, which
is $\propto \sum_{l=0}^\infty P_{2l}(x) = \frac{1}{\sqrt{8(1-x)}}
+ \frac{1}{\sqrt{8(1+x)}}$, and deducing the contribution of the
first sum from the integral (\ref{intsols}), we can write down the
solution $f_{5D}$ in closed form \cite{kalshocks}. At short
transverse distances $x = 1- H^2 {\cal R}^2/2$\, the leading order
behavior of $f_{5D}$, using $M^2_4 {\tt g} = 2M^3_5/H$ and
ignoring a constant and higher powers of ${\cal R}$, is
\begin{equation}
f_{5D}(\Omega) = - \frac{p}{2 \pi M^3_5 {\cal R}} + \frac{3p
H}{4\pi M^3_5} \ln(\frac{\cal R}{2H^{-1}})  \, . \label{5dlims}
\end{equation}
Notice the relative sign difference between the two terms on the
RHS of (\ref{5dlims}). This is necessary in order for both to
yield an {\it attractive} force $ \propto - \vec \nabla f$ on a
test particle in the shock wave background. The first term is just
the $5D$ shock wave solution of \cite{gabriele}. The second term
comes from the residual 4D graviton zero mode which persists on
the normal branch when ${\tt g} \rightarrow \infty$ because the
bulk volume remains finite. Its perturbative coupling is set by
the effective 4D Planck scale $M_{4 \, eff}^2 = 4M^3_5/3H$,
consistent with (\ref{5dlims}).
At all sub-horizon distances the inverse power of ${\cal R}$ in
(\ref{5dlims}) wins over the logarithm and so in this limit
gravity really looks five-dimensional inside the cosmological
horizon.

The shock wave solutions derived here give us a clear and explicit
demonstration that the gravitational ``filter" mechanism of
\cite{dgp} may in fact work beyond perturbation theory. The key is
that the coefficients in (\ref{facsols}) decrease with the index
$l$ of the Legendre polynomials. Since the momentum of a graviton
mode on a transverse 2-sphere along de Sitter brane is
proportional to the index of the polynomial, $q \sim H l$, this
means that the modes with $q > {\tt g} H \simeq 1/r_c$ have
amplitudes suppressed by $q$. This controls the rate of divergence
of the series, and limits it to be no worse than a logarithm for
any finite value of ${\tt g}$. Hence at short distances ${\cal R}
\le r_c \simeq H^{-1}/ {\tt g}$, gravitational shock waves are
exactly the same as in 4D GR. Only for very low momenta are the
amplitudes unsuppressed, and so the shock wave profile changes
towards 5D at large distances ${\cal R} \ge H^{-1}/ {\tt g}$. We
see this explicitly in the limit ${\tt g} \rightarrow \infty$ on
the normal branch: we recover 5D gravity since when $r_c
\rightarrow 0$ all modes with finite momenta remain unsuppressed.
Thus we expect that in DGP with finite ${\tt g}$ the leading order
Planckian scattering behaves as in 4D GR, since the differences
are suppressed by powers of $m/p$ and $H{\cal R}$, both very small
for highly energetic particles at short distances.

We note that our exact shock waves may be a new arena to explore
dynamics of the scalar graviton in DGP. Imagine probing the
gravitational field of an arbitrary mass on the brane with a very
relativistic probe. Since local physics in DGP obeys the usual 4D
diffeomorphism invariance, we can transform all of the relevant
physics to the rest frame of the probe. In this frame, the source
mass will appear to move with a very high speed ${\tt v}$, and so
its gravitational field should be well approximated by our shock
wave solutions, with the corrections due to the mass suppressed by
the powers of $m/p = \sqrt{1/{\tt v}^2-1}$. One can then treat the
rest mass of the source as a perturbation of the shock wave
geometry, and study how the scalar graviton responds to it. If
organized as an expansion in powers of $m/p$, perturbation theory
may be under control. It would be interesting to investigate what
happens with the strong coupling of
\cite{strongcouplings,lpr,giga,nira} in this case. This may shed
new light on the interplay of background curvature and scalar
graviton effective field theory.

\smallskip

{\bf \noindent Acknowledgements}

\smallskip

We thank S. Dimopoulos, G. Dvali, R. Emparan, G. Gabadadze, M.
Luty, K. Sfetsos and L. Sorbo for useful discussions, and the
Aspen Center for Physics for hospitality. NK was supported in part
by the DOE Grant DE-FG03-91ER40674, by the NSF Grant PHY-0332258
and by a RIA from the Research Corporation.

\end{document}